\documentclass[journal,comsoc]{IEEEtran}
\usepackage{graphicx,subfigure}
\usepackage[colorlinks]{hyperref}
\usepackage{amsmath,amssymb,balance}
\usepackage{booktabs}
\usepackage{array}
\newcolumntype{M}[1]{>{\centering\arraybackslash}m{#1}}
\newcolumntype{L}[1]{>{\raggedright\arraybackslash}m{#1}}
\usepackage{multirow}
\usepackage[table]{xcolor}
\newcommand{\chg}[1]{\textcolor{black}{#1}}

\begin{document}
\title{Bridging Theory and Practice in \\Reconfigurable Fluid Antenna Systems}
\author{Halvin~Yang,~\IEEEmembership{Member,~IEEE},~Yizhe~Zhao,~\IEEEmembership{Member,~IEEE},~Kai-Kit~Wong,~\IEEEmembership{Fellow,~IEEE},\\Hsiao-Hwa~Chen,~\IEEEmembership{Life Fellow,~IEEE}, and~Chan-Byoung~Chae,~\IEEEmembership{Fellow,~IEEE}

\vspace{-0.25in}
\thanks{H. Yang (email: {\tt halvin.yang@imperial.ac.uk}) is with the Department of Electrical and Electronic Engineering, Imperial College London, UK. Y. Zhao (email: {\tt yzzhao@uestc.edu.cn}) is with the School of Information and Communication Engineering, University of Electronic Science and Technology of China, China. K.-K. Wong (email: {\tt kai-kit.wong@ucl.ac.uk}) is with the Department of Electronic and Electrical Engineering, University College London, UK, and Yonsei Frontier Lab. Yonsei University, South Korea. H.-H. Chen (email: {\tt hshwchen@mail.ncku.edu.tw}) (the corresponding author) is with the Department of Engineering Science, National Cheng Kung University, Taiwan. C.-B. Chae (email: {\tt cbchae@yonsei.ac.kr}) is with the School of Integrated Technology, Yonsei University, South Korea.}}

\maketitle
	
\begin{abstract}
Fluid antennas, including those based on liquid, mechanical, and pixel-based technologies, are poised to significantly enhance next-generation wireless systems by adaptively optimizing their radiation characteristics. Many theoretical analyses assumed near-instant reconfiguration, perfect channel knowledge, static or slowly varying propagation environments, and ideal material properties that rarely hold in practice. In this article, we dissect these common assumptions and contrast them with the realities of finite actuation time, limited and imperfect channel state information, rapidly changing fading conditions, electromagnetic coupling, and mechanical constraints. Through illustrative examples and simulations, we demonstrate how ignoring these factors can lead to overestimated gains in capacity, coverage, etc.. We then propose modeling refinements, experimental validation methods, and emerging control algorithms that better account for real-world constraints. Our findings highlight that, while reconfigurable antennas remain highly promising for B5G/6G and Internet of things (IoT) applications, their full potential can only be realized by incorporating practical considerations into system design and performance evaluation. 
\end{abstract}
\begin{IEEEkeywords}
Fluid antenna, fluid antenna system (FAS), movable antenna, modeling, resource allocation, 6G.
\end{IEEEkeywords}
	
\section{INTRODUCTION}
\IEEEPARstart{W}{ireless} networks beyond fifth-generation (5G) are required to keep up with the exponential growth in \chg{mobile data traffic while also providing ubiquitous connectivity}. To achieve this, new wireless technologies need to be developed. In recent years, multiple-input multiple-output (MIMO) technology has been viewed as a primary solution to drastically improve spectral efficiency (SE) and energy efficiency (EE) of wireless communications via space-time coding, spatial multiplexing and beamforming. However, this performance improvement comes at a high cost, namely the addition of more radio-frequency (RF) chains to support \chg{additional antennas and higher computational complexity} when tackling more complex channels. The increasing system complexity, hardware size and energy consumption pose significant limitation when considering mobile or edge \chg{devices, which lack the computational capacity and energy supply of base stations} (BS). 
	
With recent breakthroughs in reconfigurable antenna technology, it is possible to design an antenna that can switch its position in a fixed space along predetermined points, each referred to as a port~\cite{ref2}. Owing to this unique ability to change between multiple spatial locations, this article refers to such a communication system as a fluid antenna system (FAS). Note that a physical antenna used in FAS is not necessarily liquid in nature; the term \emph{fluid} is a description of the smooth or dynamic nature of the antenna rather than a physical description. FAS introduces the novel concept of a position-reconfigurable antenna that can access the null of the interference created by natural fading phenomenon in a multipath-rich environment, significantly reducing the computational complexity of the system as complex beamforming and channel estimation for precoding at the transmitter are no longer required \cite{Wong-2022}.

The concept of FAS was initially introduced by Wong~{\em et~al.}~in \cite{ref2}, sparking further interest in utilizing reconfigurable antenna technologies for wireless communication system design. Following \cite{ref2}, Zhu {\em et~al.} introduced movable antenna in \cite{ref4} to describe a specific class of FAS that can physically relocate the antenna within a confined space using mechanically movable radiating elements. Recent studies even suggested 3-D positioning and 3-D orientation of the antenna surface for a 6-D movable antenna architecture. In this article, we adopt the term FAS in a broad sense to encompass various forms of spatially reconfigurable antenna systems, including fluidic, mechanically movable, and other implementations.
	
Recent advances have further envisioned to scale FAS into large reconfigurable surfaces, effectively engineering a wireless environment itself \cite{Wong-swc2021}. This concept proposes that entire building facades or urban infrastructure could be transformed into massive FAS deployments. By dynamically shaping signal paths, these enormous FAS could maximize coverage and enhance capacity in real time, extending the principles of reconfigurable intelligent surfaces (RIS) into fully adaptive communication environments. While such a deployment introduces new challenges, it represents a substantial step towards 6G-enabled smart environments, where communication optimization is embedded directly into physical infrastructure.
	
Previous literature has shown the viability of fluid antennas deployed in communication systems with a single-antenna FAS, not only significantly outperforming a traditional fixed-position antenna system \cite{ref2}, but also achieving high multiplexing and diversity gains despite being a single antenna \cite{Wong-2022}. As FAS technology advances over the years, different systems were proposed, each having different interpretation of fluid antennas, different terminologies, physical architecture, and thus different FAS technical approaches, possibly causing some confusions. Readers are advised to refer to \cite{New-2024tut} for more information. 

\chg{Current FAS research often assumes idealized material, channel, and movement conditions, which do not hold in practice.} This article will discuss those realistic conditions and their impact on communication systems. Corresponding solutions are also recommended at the end of the article, which may serve as future research directions of this emerging and exciting new research area. 

The rest of the article is organized as follows. Section~\ref{sec:overview} summarizes the key concepts, characteristics, and advantages and disadvantages of existing FASs, followed by a fundamental illustration on how a typical FAS operates. Then, Section~\ref{sec:reality} examines the theoretical assumptions made in FAS research and compares them with practical realities, highlighting their implications on system performance. Section~\ref{sec:implication} discusses the broader impact of these assumptions on communication systems. After that, Section~\ref{sec:direction} proposes solutions and outlines future research directions. Finally, Section~\ref{sec:conclude} concludes the article.

\section{OVERVIEW ON FAS TECHNOLOGIES}\label{sec:overview}
This section provides an overview on how FASs are implemented practically by looking at different fluid antenna designs, and then discusses about the fundamental operating \chg{principles of FAS and related enabling technologies}.
	
\subsection{Physical Design}
A fluid antenna refers to any radiating structure in which software-controlled conductive or dielectric elements dynamically alter their shape, position and/or geometry to reconfigure key metrics such as operating frequency, polarization, or radiation pattern. These structures may involve fluidic elements, such as conductive liquids namely, Eutectic Gallium-Indium, Galinstan, or ionized solutions (e.g., NaCl)~\cite{ref13,Tong-2025}. Depending on the design, such fluid antennas can enable frequency or spatial reconfigurability, but exhibit slow switching speeds due to fluid inertia and often require bulky pumps or reservoirs.
	
In contrast, pixel-based antennas use electronically switched static radiating elements, e.g., PIN diodes or MEMS, to form reconfigurable patterns without mechanical movement~\cite{ref7}, enabling microsecond- or even nanosecond-level switching and supporting both shape and position reconfiguration. Similarly, metamaterial-based antennas employ tunable electromagnetic surfaces composed of reconfigurable unit cells, thus achieving agile and fully electronic beam control without moving parts and offering comparable switching speeds~\cite{liu2024}.
	
By comparison, mechanical antennas physically reposition radiating elements, often in 2D or 6D space (3D translation and 3D rotation), using motors along predefined tracks. While they may offer fine spatial resolution, they often suffer from slow actuation, big energy overhead, and wear-and-tear. Contrary to some claims, mechanical antennas are not simple to implement, requiring structural supports, motion control, and synchronization. Moreover, antenna size is not architecture-dependent, and large arrays can be realized using any technology if deployment constraints permit. Table \ref{tab:fas_table} summarizes the main fluid antenna architectures, highlighting their mechanisms, benefits, and limitations.
	
\begin{table*}[!t]
\centering
\caption{Summary of Different Fluid Antenna Architectures}
\renewcommand{\arraystretch}{1.3}
\begin{tabular}{|L{2.8cm}|L{5.2cm}|M{4.0cm}|L{4.2cm}|}
\hline
\textbf{~~~~\chg{Technology Type}} & \textbf{~~~~~~~~~~~~~~~~~~Architecture} & \textbf{\chg{Operating Principle}} & \textbf{~~~~~Advantages/Disadvantages} \\
\hline
\textbf{Pixel-Based Antenna \cite{ref7}} 
& A grid of static radiating elements is reconfigured electronically by toggling integrated switches (e.g., PIN diodes or MEMS) on/off to form different radiation patterns. No physical movement is involved, allowing ultra-fast and reliable switching. Scalability depends on the number and layout of switches.
& \includegraphics[width=0.36\linewidth]{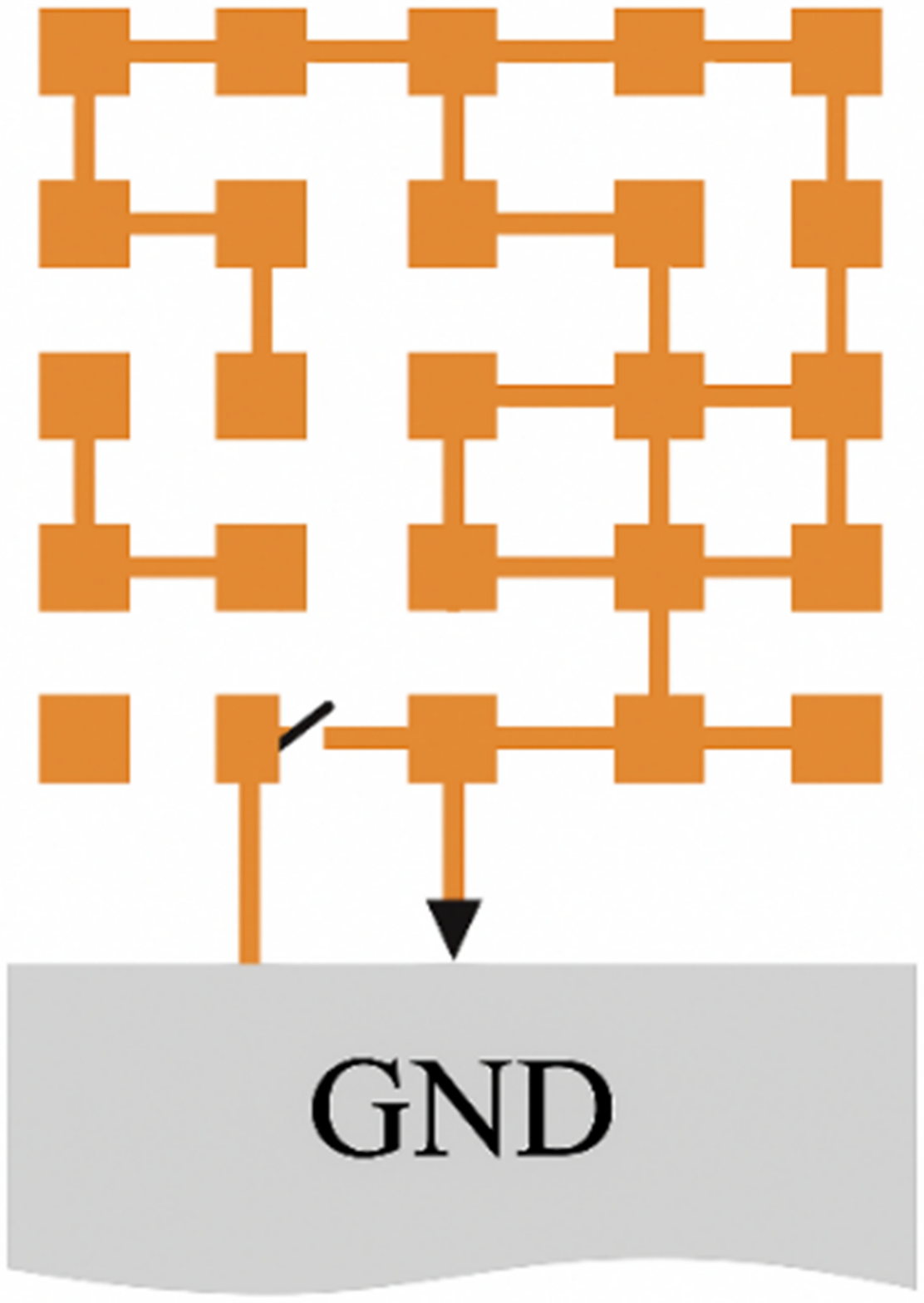}
& \begin{itemize}
\item Extremely high-speed switching (in $\mu$s range, even reaching $n$s)
\item Limited by the number of switches
\item Large unused sections
\item Design complexity increases with reconfiguration capability
\end{itemize} \\
\hline
\textbf{Mechanical Antenna \cite{ref4}} 
& Antenna element is repositioned mechanically in 2D or 3D space using motors and a structural frame with predefined paths or tracks. Advanced variants (e.g., 6DMA) support simultaneous translation and rotation. It requires physical support infrastructure, power control, and real-time coordination. Signal is routed through an RF chain for processing.
& \includegraphics[width=0.9\linewidth]{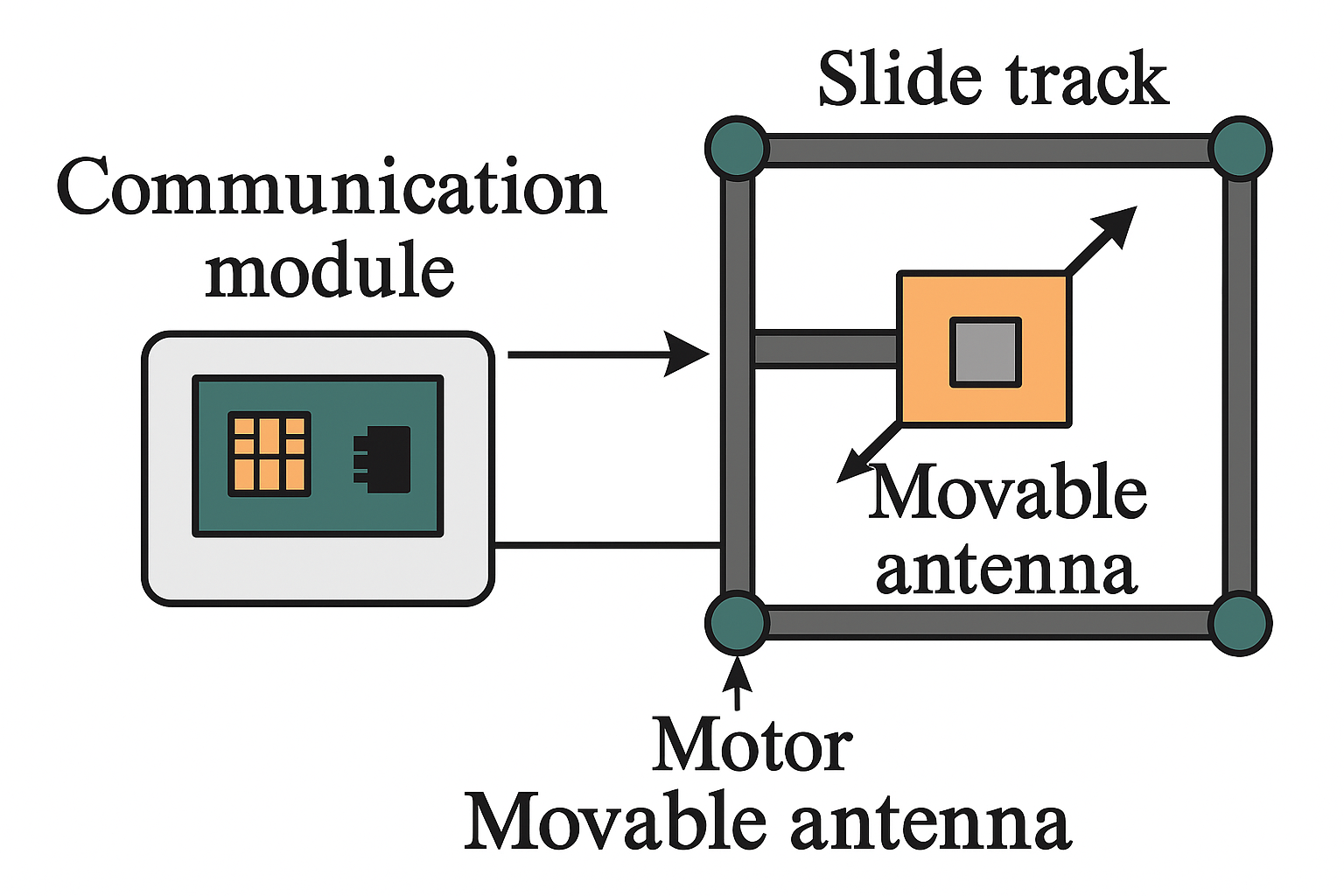}
& \begin{itemize}
\item 2D or 6D spatial coverage depending on configuration
\item Slow switching response (similar to liquid-based fluid antennas)
\item Requires structural frame, control logic, and mechanical actuation
\item Power consumption and mechanical wear were often overlooked in analysis
\item Weight and size limitations depend on deployment
\end{itemize} \\
\hline
\textbf{Surface-Wave Position-Flexible Antenna \cite{ref13,Tong-2025}}
& It combines fluid monopole and surface-wave designs under a shared category. Liquid-based antennas control RF behavior by relocating conductive fluid inside a channel. Frequency-reconfigurable designs (monopole type) vary the fluid's length and surface-wave types adjust fluid position to steer or receive guided waves. It is typically actuated using pumps or pressure gradients.
& \includegraphics[width=0.9\linewidth]{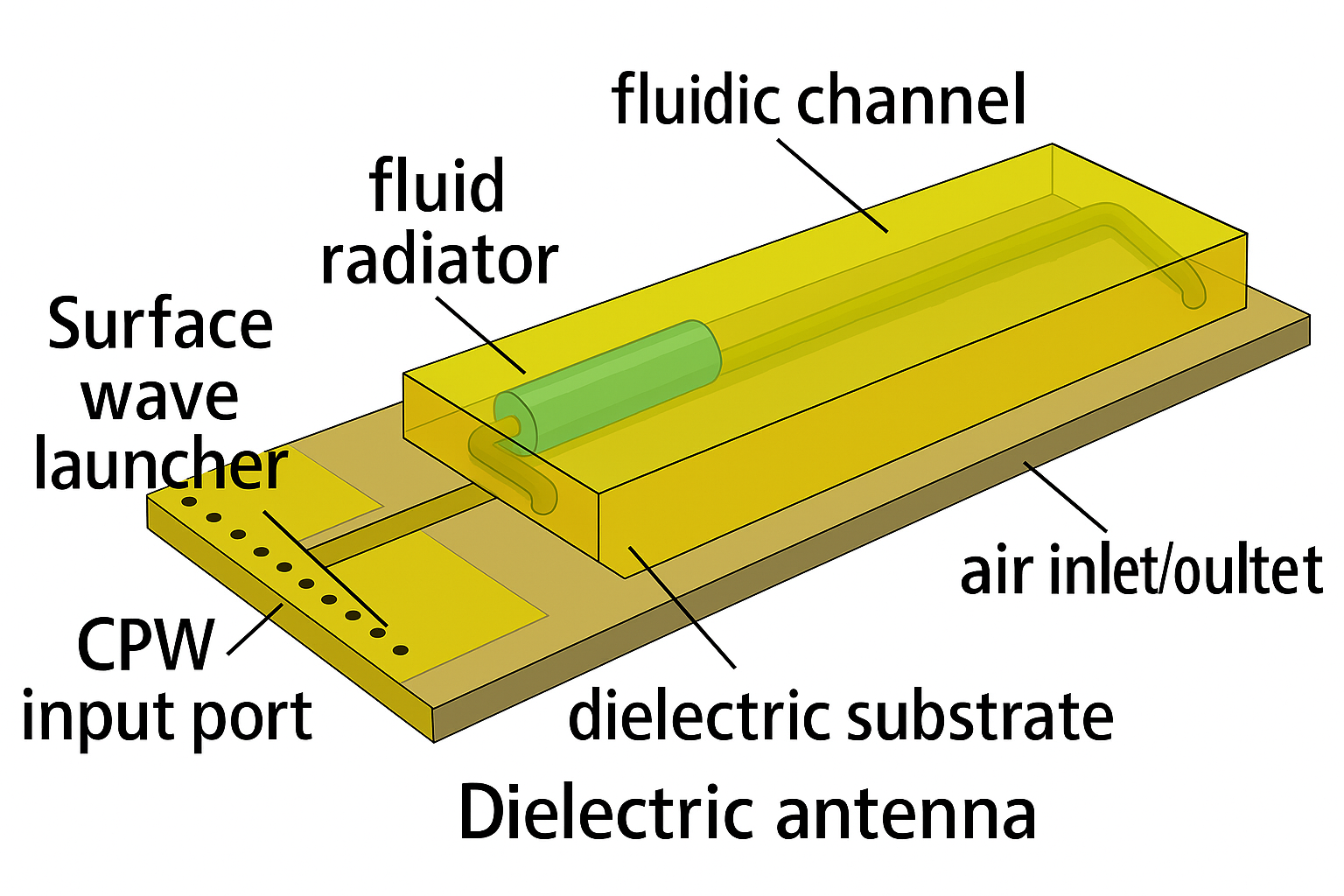}
& \begin{itemize}
\item Enables frequency or position reconfigurability depending on design
\item Moderate spatial control precision
\item Slow switching speed due to fluid inertia
\item Design may be bulky due to pumps and reservoirs
\item No solid-state switching, with limited responsiveness compared to pixel or metamaterial designs
\end{itemize} \\
\hline
\textbf{Metamaterial-Based Antenna \cite{liu2024}} 
& It utilizes programmable metasurfaces or tunable unit cells to manipulate electromagnetic wave propagation without mechanical movement. These structures can dynamically adjust their properties to control beam direction, shape and polarization, enabling agile and efficient beamforming.
& \includegraphics[width=0.9\linewidth]{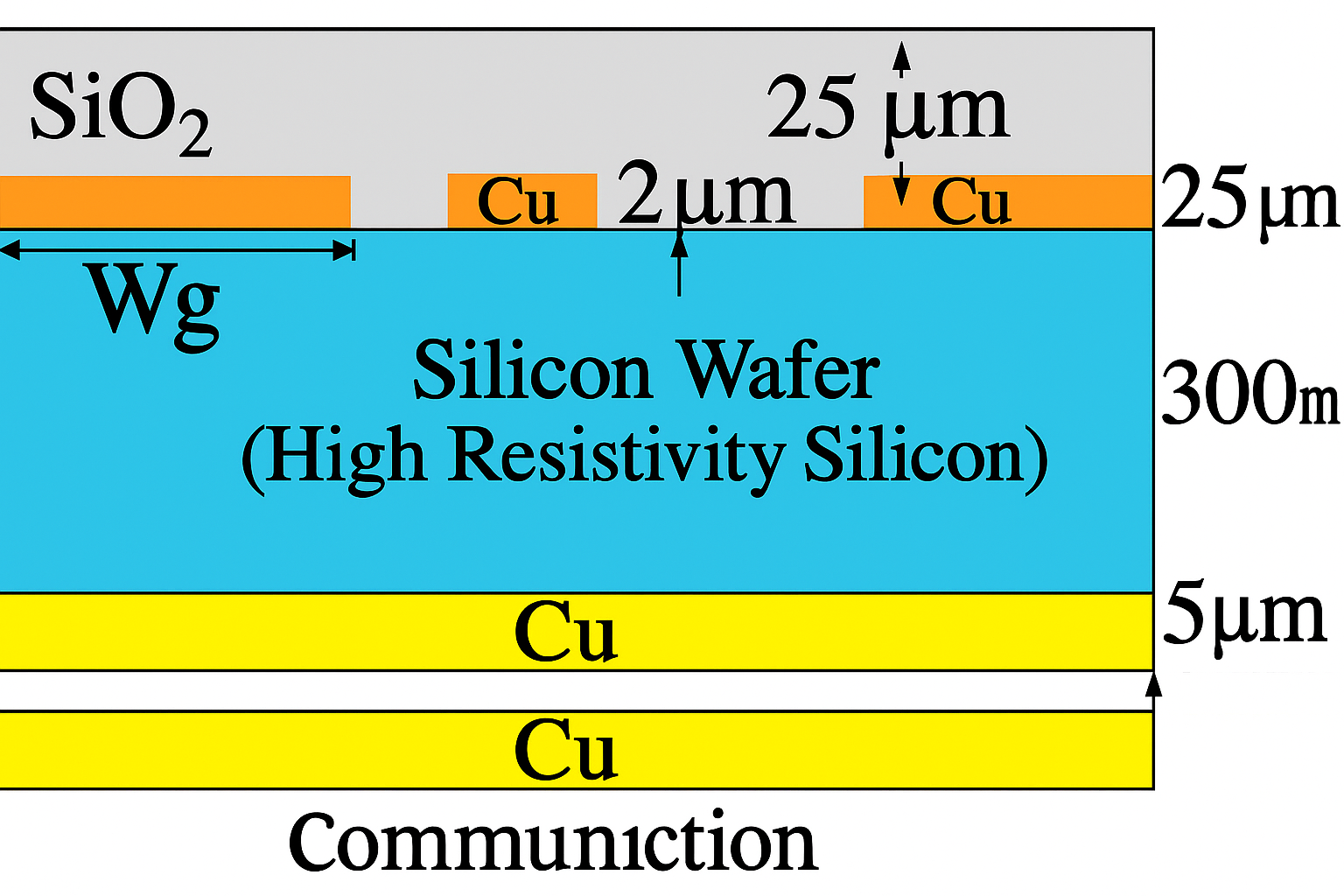}
& \begin{itemize}
\item No mechanical movement, and fully electronic reconfiguration
\item Ultra-fast switching speeds
\item Compact and integrable into various form factors
\item Requires complex control circuitry
\item High design and fabrication complexity
\end{itemize} \\
\hline	
\end{tabular}
\label{tab:fas_table}
\end{table*}

\subsection{Operating Principles}
\begin{figure}[t]
\centering
\includegraphics[width=0.45\textwidth]{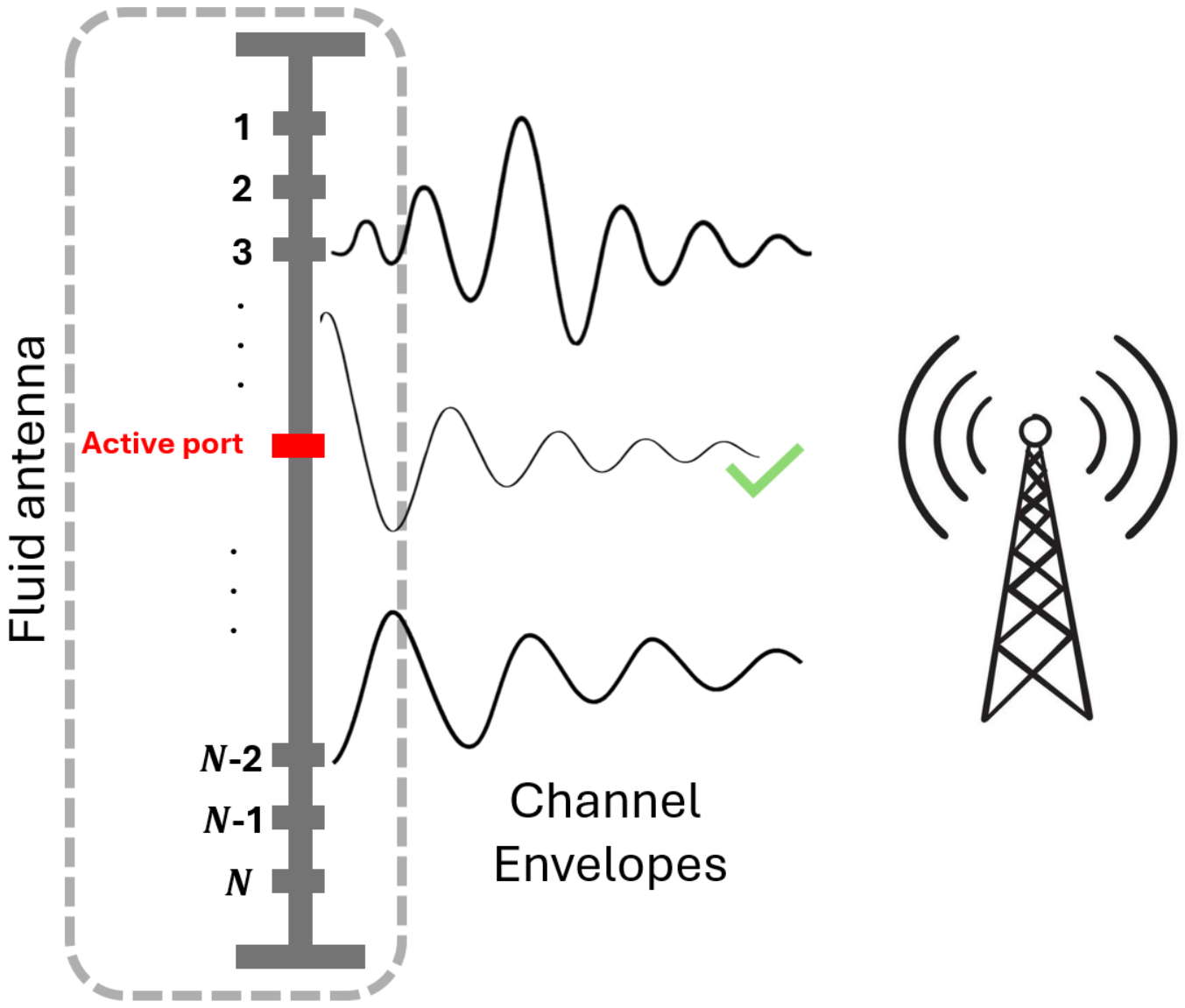}
\caption{An illustration of a FAS's browsing capabilities~\cite{ref2}.}\label{fig:FAS_browsing}
\end{figure}

Fig.~\ref{fig:FAS_browsing} exemplifies a FAS. By enabling radiating elements to change positions in space, a fluid antenna allows users to browse through different fading envelopes located at each different spatial location. Not only does this enable benefits to diversity and capacity, but also can users select the position in space, where the signal power is the highest and the noise or interference is suppressed by a deep fade. 

By exploiting the capability of FAS to resolve interference by selecting the optimal port (location in space), multiuser communication can be achieved without any complex beamforming or channel feedback. In a traditional MIMO system, where time-frequency resources are shared, signal processing and beamforming were required to create peaks and nulls of the fading envelopes with channel state information (CSI) beforehand. The interference at each user in a FAS is expected to be mitigated by the fluid antenna via a process known as port selection, in which the port with the highest signal-to-interference plus noise ratio (SINR) can be selected. A more sophisticated scheme that selects more than one ports is also possible to enhance interference immunity \cite[Section V-D]{New-2024tut}.

\subsection{State of the Art}
\begin{itemize}[]
\item \textbf{Channel Characteristics:} Due to the proximity of ports, which can be arbitrarily close, correlation between ports is a key consideration in FAS. Thus, to get an accurate representation of a FAS channel, the correlation needs to be modeled accurately because an underestimation of the correlation would result in unrealistic performance evaluation. There are already several correlation models proposed recently, including the block correlation model, which assumes a large number of correlated random variables expressed in a matrix, and the single port correlation model, in which it is assumed each port is correlated with a single reference port and therefore there is one mutual random variable at each port. The block correlation model exhibits the highest accuracy, while single port correlation model has a lower analysis complexity \cite[Section II]{New-2024tut}.
\item \textbf{Modeling and Performance Analysis:}  Performance analysis is essential in understanding the viability of FAS in a communications environment. To this end, a majority of previous works was focused on the derivation of such performance indicators as outage probability, multiplexing gain, bit error rate, and data rate. The accuracy of the channel model will impact the performance analysis on different indicators in different scenarios, while more accurate models tend to be significantly more complex. 
\item \textbf{Resource Allocation and Optimization:} \chg{Resource allocation and optimization strategies also depend on the underlying FAS architecture. Pixel-based and metamaterial antennas, due to their ultra-fast switching, are particularly suited for optimization problems that require frequent reconfiguration, such as fast port selection, user scheduling, and multiuser interference suppression. By contrast, mechanical and liquid-based designs face slower actuation and higher energy costs, making them less effective for real-time optimization but still valuable in long-term tasks such as static port assignment, spectrum planning, or energy efficiency optimization. Liquid and surface-wave designs further enable frequency and position agility, which can be exploited in spectrum allocation and interference avoidance, albeit with slower responsiveness. Hence, the choice of optimization strategy should reflect both the communication objective and the physical limitations of the chosen architecture.}

%Port selection is considered as a way to optimize the spatial resource, which is the process of identifying the optimal port in a specific communication environment. This normally comes in the form of the port, where the interference is at a null while the signal power is the highest, or in other words the port with the highest SINR is selected. Apart from port selection, other optimizations are also necessary when considering multiuser scenarios. For instance, traditional resources, such as transmit power, bandwidth, time slots or beamforming vector, can be dynamically allocated by integrating with port selection for the sake of improving communication performance of the whole network.
\end{itemize}

\begin{figure*}[t]
\centering
\includegraphics[width=0.6\textwidth]{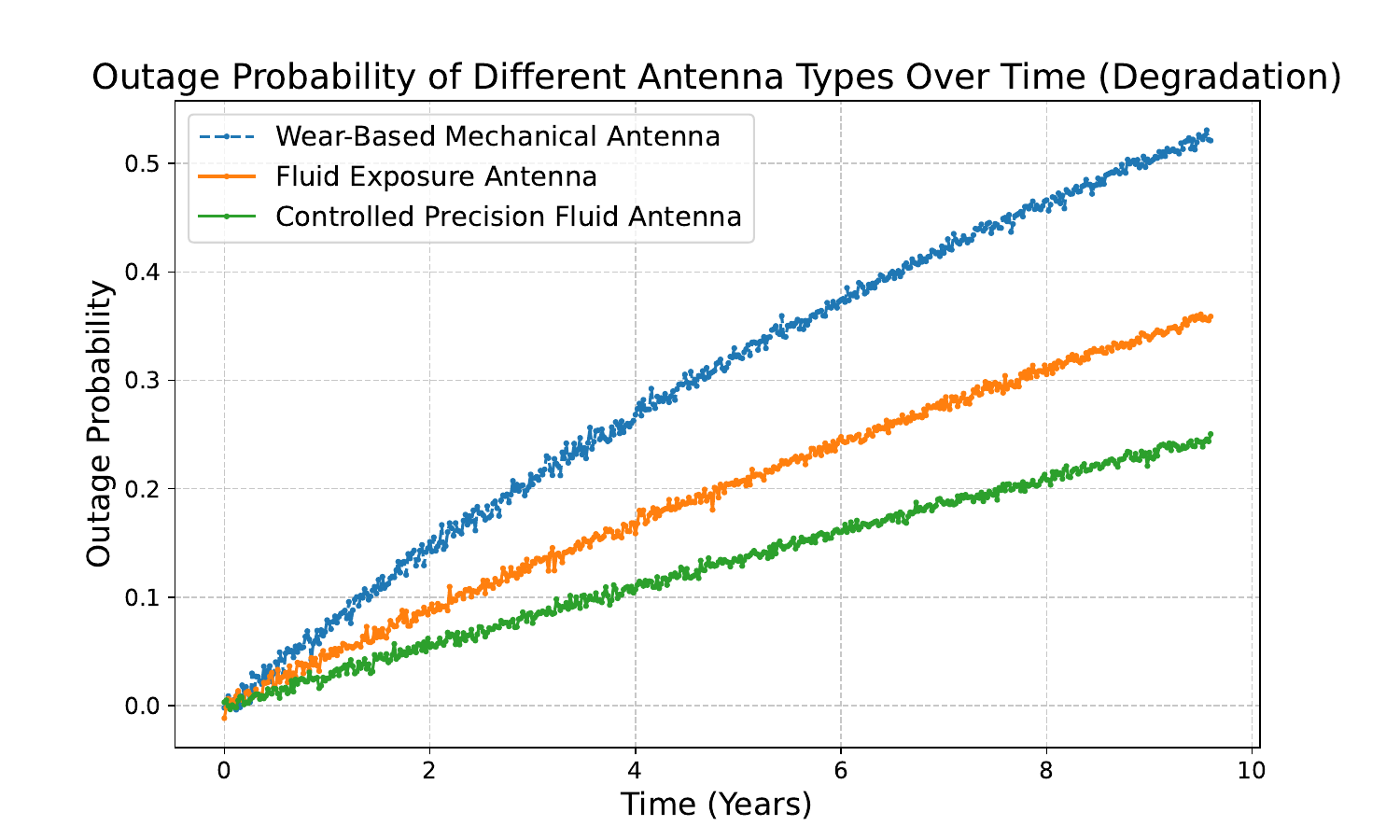} 
\caption{\chg{Performance degradation of three antenna types over time. Outage probability trends were simulated with gradual signal loss and small random variations to reflect environmental impacts such as wear-and-tear, fluid exposure, and temperature changes.
}}\label{fig:performance_degradation}
\end{figure*}

\section{THEORETICAL ASSUMPTIONS VS REALITY}\label{sec:reality}
Fluid antennas, often described as reconfigurable, mechanical, or pixel-based solutions, promise to dynamically modify antenna parameters to maximize signal-to-noise ratio (SNR), data rates, or coverage. Despite this potential, many studies relied on simplified assumptions that may lead to inflated expectations or overly idealized system models. In the text followed, we examine these assumptions and contrast them with practical constraints. We also discuss the implications of these idealized assumptions on communication system performance.

\subsection{Ideal Material Properties}
\subsubsection*{Assumption} Existing fluid-based antenna models treat the fluid as having constant, frequency-independent conductivity and negligible temperature dependence. Similarly, structural materials may be taken as perfect dielectrics or perfectly flexiblity. Analytical models often assumed that once an optimal configuration is found, the antenna can reliably return to that exact state with minimal error each time.

\subsubsection*{Reality} Two issues need to be considered carefully:
\begin{enumerate}
\item Frequency Dependence: Conductive liquids, gels, and alloys often exhibit dispersive properties, where conductivity and permittivity vary with frequency. 
\item Thermal and Aging Effects: Temperature shifts can alter fluid viscosity or cause metal oxidation, leading to performance drifts. Over time, repeated mechanical or chemical stress can degrade materials as well \cite{kubo2010}.
\end{enumerate}

\subsubsection*{Impact} Ignoring real materials' frequency dependence and aging effect leads to overly optimistic predictions of antenna gain and efficiency. Designers may find their prototypes underperforming in practical temperature ranges or long-term deployment. More experimental data on these issues can be found in \cite{kubo2010}. In practice, a `best configuration' might vary slightly each time. This inconsistency impacts link reliability and may require additional measurements or dynamic calibration, increasing overhead. This may require \chg{calibration loops to ensure performance reliability} \cite{kubo2010}. Fig.~\ref{fig:performance_degradation} shows a rough indication of the impact of wear-and-tear on the performance of antennas, showing the impact of antenna wear due to mechanical movement, antenna fluid degradation from environmental exposure, and the impact of degradation of control mechanisms.

\begin{figure*}[t]
\centering	
\subfigure[]{
\includegraphics[width=0.6\textwidth]{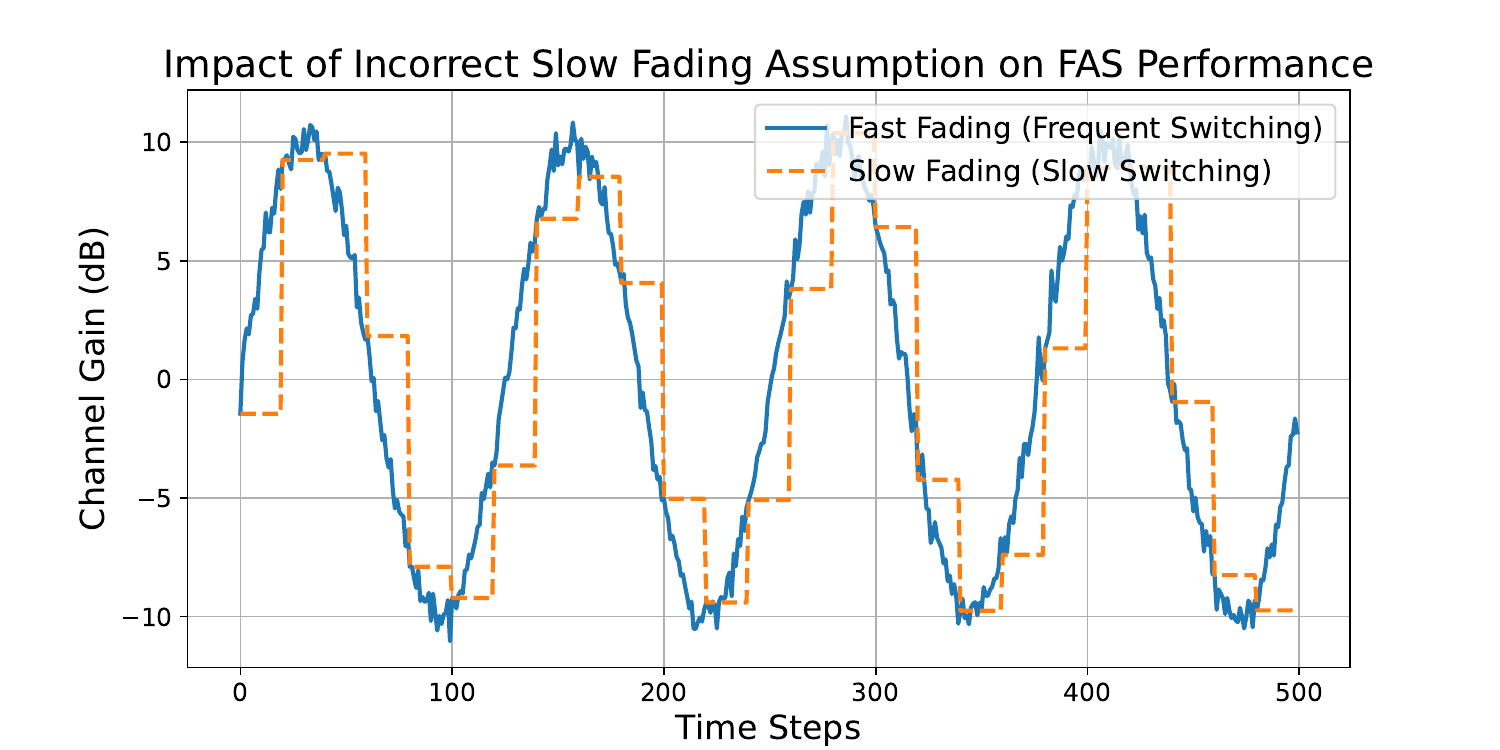}}
\vspace{0.5cm}	
\subfigure[]{
\includegraphics[width=0.6\textwidth]{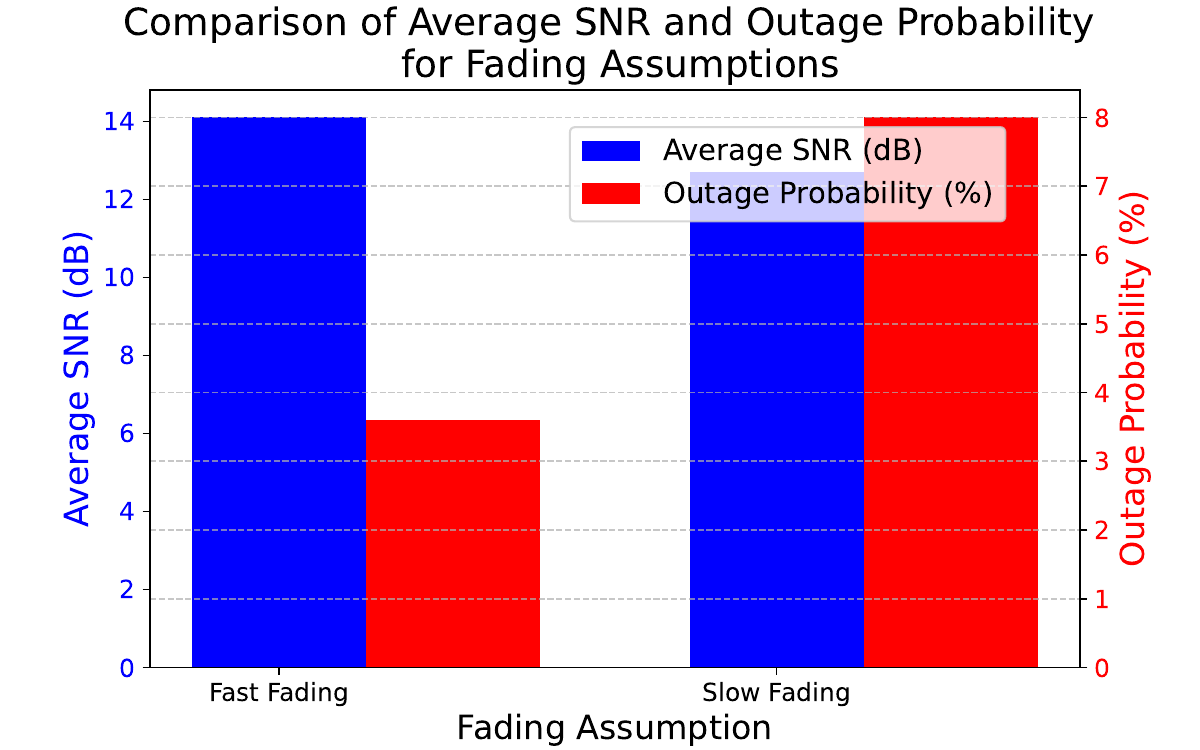}}
\caption{The effect of an inaccurate fading assumption on the performance of the system: (a) the channel gain of slow and fast fading assumptions in a fast fading environment, and (b) the SNR and outage performance of fading assumptions in a fast fading channel. \chg{These results illustrate the importance of modeling realistic channel dynamics for FAS.}}\label{fig:fading_comparision}
\end{figure*}

Fig.~\ref{fig:performance_degradation} illustrates how three different antenna types degrade in performance over a ten-year period, expressed in terms of an increasing outage probability, the likelihood that the antenna fails to maintain a reliable connection. Each curve was generated by simulating a gradual decline in signal quality over time, with added random variation to reflect environmental impacts like weather, dust, or material inconsistencies. The mechanical antenna assumes a steady performance loss from physical wear, corrosion, and moving-part fatigue. The fluid exposure antenna degrades more slowly, capturing the effects of fluid leakage or contamination over extended use. The directional fluid antenna shows the most stable behavior, thereby reflecting its reliance on solid-state control components rather than moving parts or exposed fluids. The rate at which performance degrades varies across the antennas, depending on how sensitive each design is to material or control deterioration.

\subsection{No Impact of Packaging or Enclosures}
\subsubsection*{Assumption} Antenna performance simulations were usually performed in free space, assuming no physical constraints on movement or additional losses from device enclosures.

\subsubsection*{Reality} The actual performance does depend on the following factors.
\begin{enumerate}
\item Space Constraints: Mobile handsets, IoT devices, or unmanned aerial vehicles (UAVs) have different limiting factors that restrict antenna motion or fluid distribution pathways \cite{ref13,Tong-2025}. 
\item Interaction with Enclosures and Users: Plastics, metals, and even user's hands can detune the antennas, causing modeled reflections and absorption.
\end{enumerate}

\subsubsection*{Impact} In practice, antenna elements often cannot be physically moved into theoretically `optimal' positions. Real packaging might reduce the reconfigurable range, adding new losses or scattering effects that degrade performance gains.

\subsection{Negligible Mechanical and Electromagnetic Coupling}
\subsubsection*{Assumption} Some studies analyzed a single `element' or assumed perfect isolation between antenna elements, ignoring the mechanical constraints or electromagnetic interactions that naturally arise in closely packed arrays. The change in antenna geometry in any form can have a significant impact on the RF performance of the whole system, or in the case of multiple antennas affect adjacent radiating elements \cite{zhang2021}. 

\subsubsection*{Reality} The following effects matter.
\begin{enumerate}
\item Mutual Coupling: In multi-element reconfigurable systems, shifting or resizing one antenna can detune its neighbors, altering the collective radiation pattern in unpredictable ways \cite{zhang2021}. 
\item Mechanical Friction and Alignment: Real mechanical actuators encounter gear backlash, friction, and alignment tolerances, meaning that the intended final position often differs from the actual one. 
\item Actuator Precision: Mechanical and fluidic systems each have tolerances, and small deviations may accumulate over multiple cycles, which could become unacceptable.
\end{enumerate}

\subsubsection*{Impact} Overlooking these coupling effects can invalidate the assumed radiation pattern, mismatch, and overall system performance. Reconfiguration might improve one metric (e.g., gain in a certain direction) but degrade the others (e.g., interference to neighboring elements). 

\subsection{Instantaneous Reconfiguration}
\subsubsection*{Assumption}
Many analyses assumed that antennas can be reconfigured with negligible latency and without additional energy overhead. Under such assumptions, the system can always `hop' to the best configuration whenever channel conditions change. However, this is often unrealistic across multiple antenna types. For instance, \cite{ref13,Tong-2025} showed that liquid-based antennas require pumps or valves and involve non-negligible delays. Similarly, mechanical antennas must physically reposition the radiating elements using motors, which also incurs latency and energy cost. Even electronically reconfigurable designs, such as pixel or metamaterial-based antennas, can involve control signaling, timing constraints, or driver delays.

\subsubsection*{Reality} The following factors must be considered.
\begin{enumerate}
\item Latency and Energy: Actuation hardware, no matter whether involving fluid redistribution, motorized movement, or complex switching logic, consumes both time and power. Frequent reconfiguration can drain battery life in portable devices and create communication gaps during the transition.  
\item Control Overhead: Synchronizing reconfiguration commands with fast-changing channels requires additional processing and signaling, introducing further delays in system responsiveness.
\end{enumerate}

\subsubsection*{Impact} Ignoring reconfiguration latency risks exaggerating theoretical throughput or coverage gains. In practice, systems may lose critical time adapting to channel conditions that might have already changed. Scheduling algorithms that assume instantaneous reconfiguration often overlook switching delays or transitional downtimes, leading to overly optimistic performance predictions that are difficult to reproduce in real deployments. Low-latency or mission-critical applications, such as industrial IoT and autonomous systems, are particularly sensitive to such overhead. Latency introduced by antenna switching may violate strict quality-of-service (QoS) requirements. Moreover, misaligned timing or rapid channel variation can degrade SNR and compromise system reliability if reconfiguration is not executed with a sufficient precision.

\subsection{Static or Slowly Varying Channels}
\subsubsection*{Assumption} Another common assumption is that the channel remains largely unchanged during the reconfiguration process, or it only changes really slowly that once the `best' antenna state is found, or it remains optimal for long periods \cite{zou2024}. This assumption may not be realistic, as \cite{gupta2018} showed that millimeter-wave channels can fluctuate rapidly. 

\subsubsection*{Reality} In real-world conditions, we have to take into account the following factors.
\begin{enumerate}
\item Mobility and Fast Fading: In urban or vehicular settings, multipath components vary rapidly. By the time an antenna arrives at its new state, the optimal direction or frequency response may have shifted.
\item Short Coherence Times: In high-frequency bands (e.g., millimeter-wave or terahertz bands), coherence time can be on the order of milliseconds, challenging any reconfiguration that takes longer time than that. 
\item Channel Dynamics: Mobility and environmental changes can cause rapid channel fluctuations.
 \end{enumerate}
 
\subsubsection*{Impact} Models assuming quasi-static channels can largely overestimate achievable performance in real dynamic scenarios. Protocols that rely on repeated reconfiguration may struggle to keep pace with channel variations, leading to underutilized potential or wasted energy. The work in \cite{gupta2018} offered further insight into the performance of FAS specifically under these fast channel variations and the detriment to performance from assuming a slow fading channel. Fig.~\ref{fig:fading_comparision} illustrates the impact of inaccurate fading assumptions. These inaccuracies and a larger channel frequency fluctuations may result in the optimal port having to change more frequently, requiring more processing power for port selection. \chg{It is worth noting that the results in Fig.~\ref{fig:fading_comparision} are based on a generic FAS channel model rather than a specific hardware implementation. The intention is to illustrate the universal impact of inaccurate fading assumptions, which all FAS architectures are subject to. Nevertheless, the severity of degradation can vary: pixel-based and metamaterial antennas, with their fast electronic switching, can adapt more effectively to fast-fading environments, whereas mechanical and liquid-based designs are more susceptible to performance loss due to their slower actuation.}

\subsection{Perfect Knowledge of Channel }
\subsubsection*{Assumption} It is often assumed that the transmitter (or a central controller) has perfect knowledge of the instantaneous CSI for all possible antenna states, enabling optimal selection among them. This is a very strong assumption, as shown in \cite{wang2024}, which highlighted the measurement complexity in capturing channel responses for numerous possible antenna states. 

\subsubsection*{Reality} Acquiring CSI comes with practical issues.
\begin{enumerate}
\item Measurement Overhead: Probing all candidate antenna states to obtain CSI can be prohibitively time-consuming and complex, especially in dense multipath or high-frequency (millimeter-wave) bands.
\item Estimation Errors: Noise, limited training symbols, and pilot contamination (in multiuser scenarios) may degrade CSI estimation accuracy.
\end{enumerate}

\subsubsection*{Impact} 
\begin{figure*}[]
\centering
\includegraphics[width=0.60\textwidth]{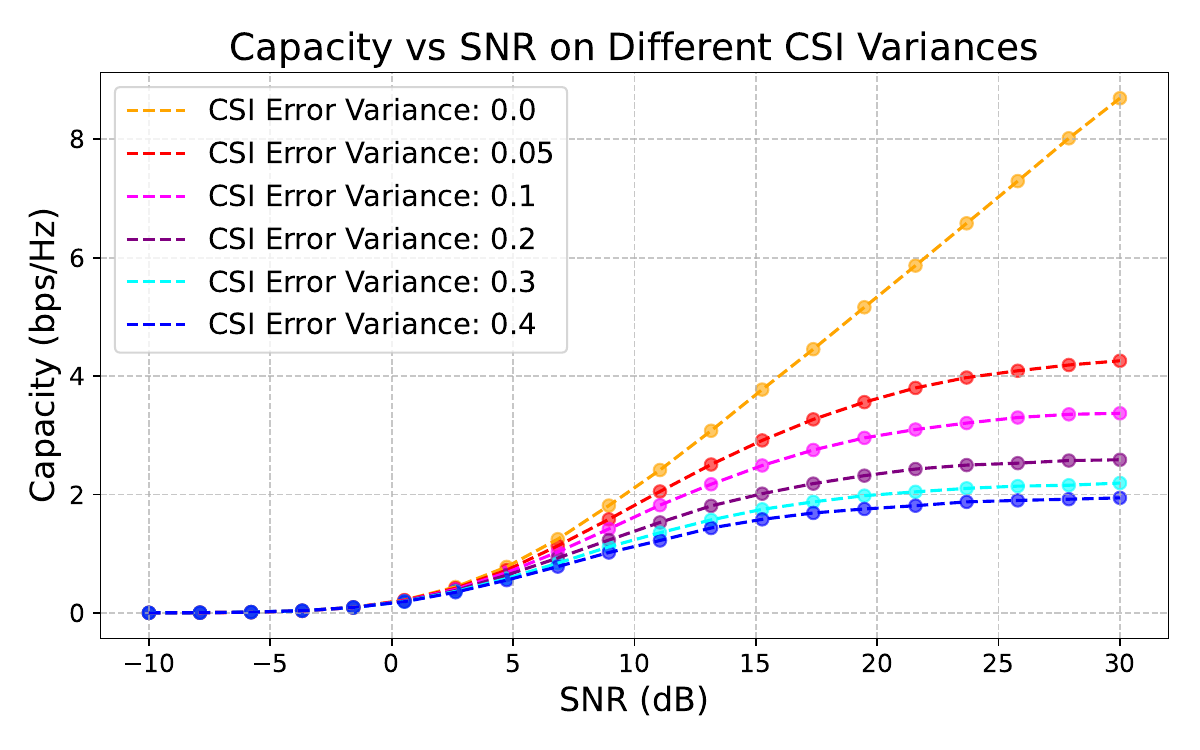}
\caption{\chg{Impact of imperfect CSI on FAS capacity. Simulations used a flat Rayleigh fading channel with Gaussian signaling and CSI errors modeled as Gaussian noise. Higher error variance significantly reduces capacity at a high SNR.}}\label{fig:csi_error}
\end{figure*}

The mismatch between assumed perfect CSI and real, noisy, partial, or outdated CSI can greatly reduce the system's ability to consistently select the best antenna configuration. Ultimately, capacity gains predicted by idealized models often shrink significantly when estimation errors or feedback delays are included. The challenges of CSI overhead were further discussed in \cite{wang2024}, which also explored the concept of adaptive selection from multiple antenna states. The impact of CSI errors on FAS can be observed in Fig.~\ref{fig:csi_error}. 

Fig.~\ref{fig:csi_error} shows the impact of imperfect CSI on wireless link capacity across an SNR range from $-10~{\rm dB}$ to $30~{\rm dB}$. The simulation assumed a flat Rayleigh fading channel with Gaussian signaling and no feedback. CSI errors were modeled as zero-mean complex Gaussian noise with variances ranging from $0$ (perfect CSI) to an error of $40\%$. Each curve represents the average of $1,000$ Monte Carlo trials. The results reveal that while CSI errors have minimal effect at a low SNR, their impact grows significantly in a high-SNR region, in which even a modest estimation error can severely limit achievable capacity. This underscores the critical role of accurate channel estimation in high-performance communication systems.

\section{IMPLICATIONS IN COMMUNICATION SYSTEMS}\label{sec:implication}
\begin{figure*}[]
\centering
\includegraphics[width=1.0\textwidth]{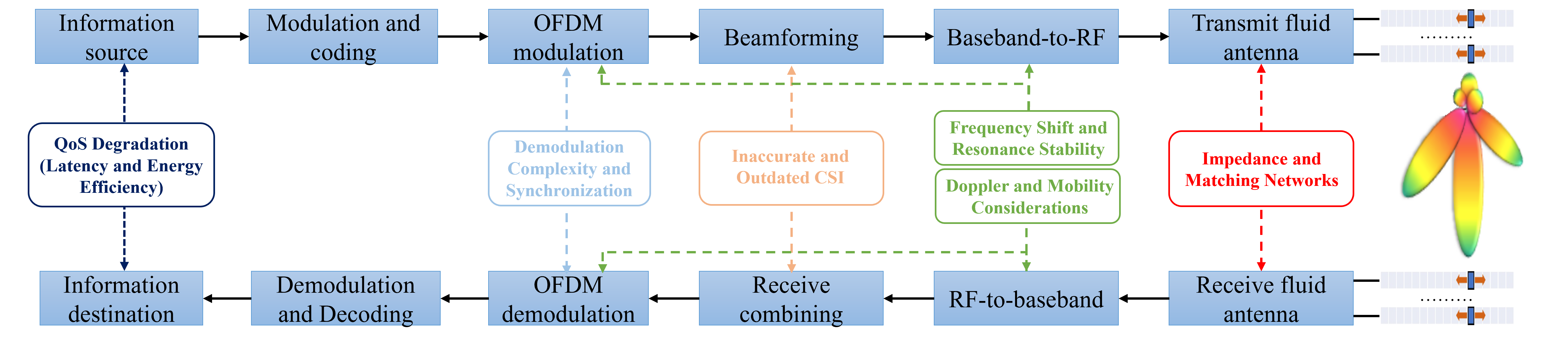}
\caption{Impacts of a non-ideal fluid antenna on communication systems.}\label{fig:impact}
\end{figure*}

The ideal assumptions discussed above can collectively lead to multiple implications on communication systems in real-world scenarios, as illustrated in Fig.~\ref{fig:impact}. Some important implications are discussed below. These physical layer impacts underscore the need for joint mechanical-electrical modeling, ensuring the dynamic antenna behavior is captured in link-level simulations. By anticipating the shifts in resonance, planning for demodulation disruptions, and incorporating offset-tracking loops, designers can mitigate transient degradations and preserve robust physical-layer performance.

\subsection{Impedance and Matching Networks} 
Fluid movement can change antenna's impedance profile, impacting the match to transmission line or power amplifier \cite{ref14}. A mismatch leads to reflected power and potential amplifier distortion, necessitating either real-time tuning (e.g., via variable matching networks) or the use of conservative design margins that reduce the benefits from reconfiguration.

\subsection{Frequency Shift and Resonance Stability} 
Fluidic antennas typically rely on shape or volume changes to alter their resonance. If the antenna's effective electrical length shifts unpredictably during or after reconfiguration, then the system can experience frequency offset and mismatches in the expected resonant band \cite{ref14}. This may cause unexpected errors for RF-to-baseband and orthogonal frequency division multiplexing (OFDM) demodulation modules, which is particularly significant for higher-frequency operation (e.g., in millimeter-wave bands), where small dimensional changes can result in large resonance shifts.

\subsection{Doppler and Mobility Considerations} 
Although fluidic antennas are not necessarily moving at high speeds themselves, rapid mechanical actuation or fluid flow can momentarily mimic Doppler-like effects in the air interface. This can lead to apparent frequency offsets during reconfiguration, requiring updated CFO (i.e., carrier frequency offset) tracking at the receiver. In high-mobility scenarios, such as vehicular communications, fluid antennas must not only cope with channel fading but also ensure that any antenna reconfiguration does not significantly worsen the already challenging Doppler spread. This may have similar impacts as the aforementioned frequency shift on a communication system.

\subsection{Inaccurate and Outdated CSI} 
The massive antenna positions make it impractical to estimate a wireless channel of all antenna states, which results in imperfect CSI at a transmitter. Meanwhile, fast fading in rapidly varying channels may occasionally exceed channel estimation frequency, which makes the acquired CSI outdated. Therefore, the beamforming efficiency can be readily degraded, since transmit precoding and receive combining both have strict requirements of perfect CSI.

\subsection{Demodulation Complexity and Synchronization} 
Real-time changes in the antenna gain pattern can introduce additional phase noise and time-varying channel responses, complicating receiver synchronization and channel estimation. For example, as the antenna reorients or reshapes, partial CSI can become obsolete much more quickly \cite{ref13}. Consequently, demodulation algorithms may need frequent pilot re-insertion or adaptive equalization to handle dynamic interference or varying signal strengths during reconfiguration events.

\subsection{QoS Degradation} 
Dynamic FAS positioning inevitably incurs additional latency and energy consumption, which further degrades the energy efficiency of a communication link. Moreover, the wasted time for antenna re-positioning also contributes to the loss of throughput. These factors altogether result in the QoS degradation of the end-to-end communication link.
	
\section{RECOMMENDATIONS AND FUTURE DIRECTIONS}\label{sec:direction}
Bridging the gap between theoretical promise and commercial deployment of reconfigurable antenna systems requires fresh perspectives and rigorous research efforts considering different aspects. Below are several key directions from the perspective of wireless communications. Innovations in other fields such as materials can also benefit FAS. For instance, there is a need for stable high-performance fluids. New conductive fluids with a lower viscosity and reduced temperature sensitivity can enhance reliability. Metallurgical advances in gallium-based alloys or ion-rich solutions can improve conductivity and longevity. Durable mechanical components are also necessary. Shape-memory alloys, flexible plastics, or 3D-printed waveguides may enable smoother and more repeatable motion, mitigating mechanical wear-and-tear issues. Apart from these, there are some promising directions that deserve more investigations, which we briefly discuss as follows.

\subsection{Stochastic or Hybrid Channel Models} 
Because channels and reconfiguration processes are inherently uncertain and time-varying, purely deterministic models are insufficient. Researchers are beginning to adopt stochastic approaches, where each antenna configuration transition is assigned a probability of success, a mean delay, and an associated CSI uncertainty distribution. Hybrid models that blend measurement-based data with theoretical frameworks can more accurately reflect practical performance envelopes, guiding realistic system-level optimizations.
	
\subsection{Medium Access Control Protocols} 
Medium access control (MAC), which relies on fine-grained on-the-fly antenna configuration changes, must incorporate latency and channel-estimation overhead into their scheduling mechanisms. Slot-based protocols, for example, may require longer guard intervals or advanced predictive algorithms if the antenna reconfiguration time is not negligible. In addition, resource allocation strategies that assume perfect CSI for all antenna states need to incorporate robust feedback loops, reducing effective spectral efficiency. In multiuser scenarios (e.g., MIMO networks), fluid antennas might coordinate with each other to conduct cooperative positioning in order to avoid interference or optimize coverage collectively.
	
\subsection{Cross-Layer Design} 
Reconfigurable antennas can no longer be viewed solely as a physical-layer phenomenon. Cross-layer frameworks, where changes at the physical layer interact with MAC scheduling and even application-layer demands, will be essential. For instance, when reconfiguration overhead is high, an application-layer decision to buffer data during reconfiguration intervals may improve overall system efficiency.
	
\subsection{Advanced Control and Optimization Methods} 
More creative and practical approaches need to be sought out in order to make FAS mature for a large scale deployment. Some of the thoughts are listed as follows.
\begin{itemize}
\item Predictive and Machine Learning Techniques: Instead of reacting to instantaneous channel measurements, systems can utilize machine learning models to anticipate channel variations based on historical data, reducing unnecessary reconfigurations.
\item Limited-Codebook Approaches: To reduce the overhead of searching a continuum of possible antenna states, designers can adopt a finite codebook of well-chosen configurations. This strategy reduces complexity and channel-measurement overhead.
\item Energy-Efficient and Time-Limited Port Selection: Practical fluid or reconfigurable antennas often have multiple ports or states, each with different power-consumption and performance trade-offs. To address energy and latency constraints, researchers have proposed time-limited port selection algorithms, which optimized port usage only within pre-defined time windows. By scheduling reconfiguration intervals more judiciously, the system avoids excessive energy draw from continuous scanning of all possible antenna states. In tandem, energy-efficient metrics (e.g., bits per joule or energy per reconfiguration) can guide selection policies, ensuring that any performance gain from reconfiguration outweighs its associated cost.
\end{itemize}

\subsection{Hardware-In-the-Loop and Field Trials} 
These include: 
\begin{itemize}
\item Realistic Prototyping: Small-scale prototypes, complete setup with actuation hardware, fluid channels or mechanical pivot arms, should be tested in realistic indoor/outdoor environments. 
\item Extended Network Trials: Multi-node testbeds (e.g., in an anechoic chamber or urban corridor) can expose the interplay of mobility, interference, and real packing constraints, the factors often absent in lab-based experiments.
\end{itemize}
	
\subsection{Standardization Efforts} 
Standardization efforts are also essential. As reconfigurable antennas mature, industry and regulatory bodies may need new guidelines for testing, compliance, and performance evaluation, similar to the existing antenna standards (e.g., ETSI, 3GPP).

\section{CONCLUSION}\label{sec:conclude}
While FAS hold great promise for dynamically optimizing performance, the assumptions listed above, often made for analytical tractability or simplicity, can lead to inflated theoretical gains. In practice, real-world constraints such as time and energy costs, material properties, limited resolution, measurement and alignment errors, dynamic channels, and multiuser interference can significantly diminish these benefits. When models overlook the factors such as reconfiguration latency and energy consumption, imperfect or limited channel knowledge, mechanical and electromagnetic coupling, non-ideal materials, packaging limitations, and misalignment issues, they risk being overly optimistic. To better predict actual performance in real-world deployments, future studies should incorporate more realistic models that account for hardware constraints, reconfiguration overhead, channel dynamics, and network-level interactions.

\chg{\subsection*{Key Takeaways}
\begin{itemize}
\item Many FAS studies relied on overly ideal assumptions (e.g., perfect CSI, instantaneous switching, ideal materials).  
\item Practical constraints such as material aging, packaging limits, and switching delays can significantly reduce the expected gains.  
\item Realistic stochastic and hardware-aware models are needed to evaluate system performance accurately.  
\item Cross-layer design is essential to manage reconfiguration overheads and fast-changing channels.  
\item Prototyping, field trials, and standardization are critical steps to translate FAS theory into practical applications.  
\end{itemize}}

\balance

\vfill
\end{document}